\documentclass[journal]{IEEEtaes}
\IEEEoverridecommandlockouts

\usepackage{color,array,amsthm}
\usepackage{graphicx}

\usepackage[utf8]{inputenc}
\usepackage{setspace}
\usepackage{amsmath,amssymb,amsfonts}
\usepackage{dirtytalk}
\usepackage{mathrsfs}
\usepackage{algorithmic}
\usepackage{textcomp}
\usepackage{xcolor}
\usepackage{longtable}
\usepackage{colortbl}

\usepackage{etoolbox}
\makeatletter
\let\labelindent\relax
\usepackage{enumitem}
\usepackage{tabu}
\usepackage{array} 

\usepackage{soul}

\patchcmd{\@makecaption}
  {\scshape}
  {}
  {}
  {}
\makeatletter
\patchcmd{\@makecaption}
  {\\}
  {.\ }
  {}
  {}
\makeatother
\def\tablename{Table}
\usepackage{adjustbox}
\usepackage{siunitx}
\usepackage{gensymb}
\usepackage{bbm}
\usepackage{algorithm} 
\usepackage{amsmath}

\usepackage{multirow}
\usepackage[english]{babel} 
\usepackage{blindtext}

\usepackage{scalerel,stackengine,amsmath}

\usepackage{scalerel}
\usepackage{stackengine,wasysym}
\usepackage{verbatim}

\def\mathcolor#1#{\@mathcolor{#1}}
\def\@mathcolor#1#2#3{%
  \protect\leavevmode
  \begingroup
    \color#1{#2}#3%
  \endgroup
}
\usepackage{csquotes}
\usepackage[backend=biber,style=ieee]{biblatex}
\usepackage{float}

\usepackage{tabularx} 
\usepackage{caption}  

\usepackage{subcaption}
\usepackage{float} 

\jvol{XX}
\jnum{XX}
\jmonth{XXXXX}
\paper{1234567}
\pubyear{2022}
\doiinfo{TAES.2022.Doi Number}

\begin{document}

\title{Optical Power Beaming in the Lunar Environment}

\author{Mohamed Naqbi}
\member{Member, IEEE}

\author{Sebastien Loranger}
\member{Member, IEEE}
\author{Gunes Karabulut Kurt}
\member{Senior Member, IEEE}
\affil{Polytechnique Montréal, Montréal, QC, Canada \\
Poly-Grames reserach Center, Dept. of Electrical Engineering}

\markboth{AUTHOR ET AL.}{SHORT ARTICLE TITLE}
\maketitle

\begin{abstract}
The increasing focus on lunar exploration requires innovative power solutions to support scientific research, mining, and habitation in the Moon’s extreme environment. Optical power beaming (OPB) has emerged as a promising alternative to conventional systems. However, the impact of lofted lunar dust (LLD) on optical transmissions remains poorly understood. This research addresses that gap by evaluating LLD-induced attenuation and optimizing OPB design for efficient power delivery over long distances. A combined theoretical and simulation-based approach is employed, utilizing the T-matrix method to model LLD attenuation and Gaussian beam theory to optimize transmission and receiver parameters. The results indicate that LLD significantly attenuates ground-to-ground optical power transmission in illuminated regions, thus making OPB more suitable in darker areas, such as permanently shadowed regions or during the lunar night. Furthermore, we demonstrate that OPB can operate over long distances on the Moon while maintaining reasonable aperture sizes through appropriate optical design optimizations. These findings highlight the potential of OPB as a reliable power solution for sustainable lunar exploration and habitation.

\end{abstract}


\section{Introduction}

The renewed global focus on lunar exploration, driven by both governmental space agencies and private enterprises, emphasizes the growing ambition to establish sustainable infrastructure on the Moon. This is exemplified by initiatives such as NASA's Artemis program, which aims to return humans to the lunar surface and develop a long-term presence \cite{lin_return_2024}. A critical concern for these future lunar operations, especially in their early phases, is the need for flexible, reliable, and efficient energy transmission systems, capable of delivering power anywhere it is needed, in the Moon's harsh and dynamic environment.\\

Indeed, for a lot of cases, it is preferable to deliver energy from external sources rather than integrating power generation directly into the system. And while electrical cables are reliable on Earth, they become highly impractical on the Moon due to their significant mass and deployment difficulties. Each kilogram of material sent to the Moon requires a large amount of fuel for both launch and landing, which greatly increases the overall cost of the mission. Additionally, placing cables on the Moon's rough and dusty surface introduces further practical difficulties, especially regarding maintenance and durability, partly due to the abrasive nature of lunar dust \cite{marcinkowski_lunar_2023}.\\

To address these challenges, wireless power transmission (WPT) emerges as a promising solution, especially where an unobstructed line-of-sight is possible. The two types of WPT emerging for this type of application are microwave and optical, which differ significantly due to their distinct wavelengths. Microwave power transmission requires bulky hardware and large antennas, making it less suitable for scalable lunar operations \cite{stavnes_analysis_1992}. In contrast, the use of optical power beaming (OPB) allows for more lightweight and compact equipment \cite{gage_laser_2011}. For this reason, it emerges as a better solution. Furthermore, one should also mention that the use of such technology offers several advantages in the lunar context, compared to Earth-based systems. Indeed, the absence of an atmosphere eliminates issues such as turbulence and scintillation, atmospheric absorption and scattering, as well as weather related disruptions, all of which are significant challenges on Earth.\\

However, implementing OPB is not without its challenges. The main challenge that will be discussed is related to the presence of lunar dust. First, because it can adhere to the surfaces of equipment, including optical components, potentially blocking or damaging them. This problem has been addressed, and solutions like electrodynamic dust shields have been proposed to help keep critical surfaces free of dust \cite{tisdale_design_nodate}. Secondly, it can loft into the lunar exosphere and potentially disrupt laser transmission through mechanisms of scattering and absorption \cite{horanyi_lunar_2024}. This phenomenon arises from electrostatic processes and micrometeoroid impacts, which are unique to the lunar environment \cite{zelenyi_dusty_2020, naqbi_impact_2024, szalay_impact_2019}. During the lunar day, tiny particles of lunar dust become positively charged through photoelectric effects from sunlight. At night, they are negatively charged because of bombardment by energetic electrons in the surrounding plasma. Additionally, micrometeoroid impacts provide sufficient kinetic energy to eject particles from the lunar surface into the exosphere. These combined processes, coupled with the Moon's low gravity, facilitate the lofting of dust particles into the lunar exosphere \cite{colwell_lunar_2007, szalay_impact_2019}.\\

\begin{figure*}[ht!]
    \centering
    \includegraphics[width=\textwidth]{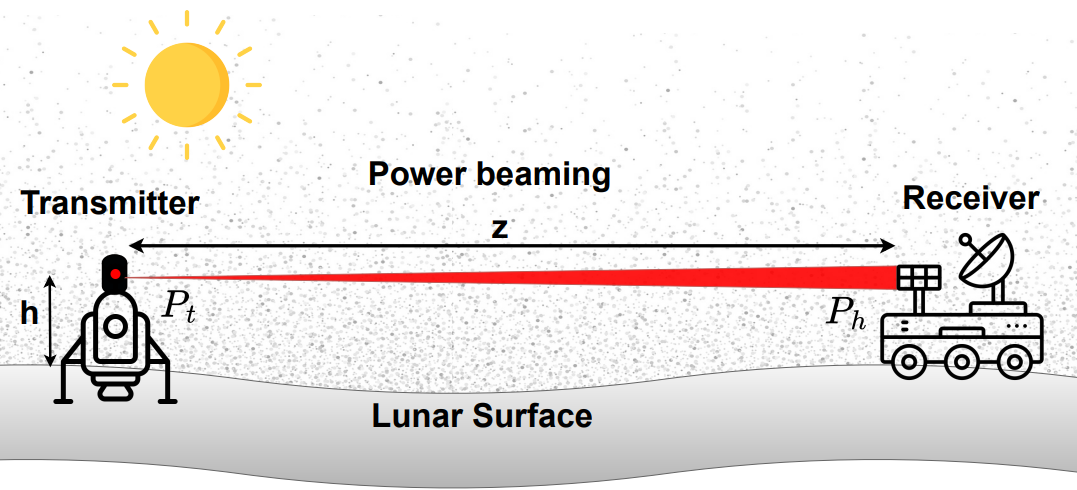}
    \caption{Ground-to-ground OPB on the lunar surface. In the Figure, $h$ represents the height of both the transmitter and receiver above the lunar surface and $z$ denotes the transmission distance.}
    \label{fig:scenario1}
\end{figure*}

\subsection{Literature Review}

Research on OPB for space applications has evolved significantly over the last decades, demonstrating the ability to wirelessly deliver electrical power over long distances. Early studies demonstrated that photovoltaic cells could achieve remarkable efficiencies when optimized for specific laser wavelengths, with efficiencies exceeding 50\% under controlled conditions \cite{olsen1991, vanriesen2002}. Over time, improvements in OPB technology have been driven by advances in laser systems and by government-led initiatives such as those from NASA, DARPA, and JAXA, along with industry collaborations with emerging companies like Powerlight Technologies. Numerous practical demonstrations, including NASA’s laser-powered aircraft experiment \cite{raible2008}, the Space Elevator Power Beaming Challenge \cite{NASA_LaserMotive_2009}, and experiments by the Naval Research Laboratory \cite{NRL2019}, have validated the feasibility of transmitting power over distances ranging from tens to hundreds of meters. These efforts have firmly established OPB as a promising technology for various applications, particularly in extreme or remote environments.\\

Despite these advancements, deploying OPB on the Moon presents unique challenges. The lunar surface is covered with fine dust particles that can become lofted above the ground due to electrostatic charging or micrometeoroid impacts \cite{colwell_lunar_2007, szalay_impact_2019}. Early observations during the Apollo missions suggested that high-concentration dust could rise to altitudes of several kilometers \cite{rennilson1974surveyor, mccoy1974lunar}. However, more recent data from the Lunar Dust Experiment (LDEX) aboard the LADEE mission have shown that the concentration is not as high as previously estimated and that electrostatic lofting is predominantly a near-surface phenomenon, while high-altitude dust is primarily the result of micrometeoroid collisions \cite{horanyi2015dust}. Although the density of lofted dust diminishes with altitude, it remains a critical concern for OPB on the Moon, particularly for ground-to-ground links on the lunar surface. Scattering and absorption of laser beams by these particles could potentially degrade the transmitted power, posing challenges to the efficiency and reliability of power delivery systems in lunar environments. This paper primarily investigates this gap by evaluating the power attenuation caused by lofted lunar dust. The second gap addressed concerns optical optimization to maximize transmission distance.

\subsection{Contributions}

This research makes the following contributions to the field of OPB in the context of a lunar scenario : \\

\begin{itemize}
    \item  
    A comprehensive evaluation of the attenuation effects caused by electrostatically lofted lunar dust (LLD) on optical power transmission is conducted. This includes the use of radiative transfer theory and the T-matrix method to quantitatively assess scattering and absorption effects, considering realistic lunar environmental conditions.
    \\
    \item
    The OPB system is optimized to maximize transmission distance while minimizing the required aperture sizes and ensuring efficient power capture. For this purpose, Gaussian beam optics is employed.
    \\
    \item   
    The relevance of using OPB on the Moon is critically analyzed by incorporating up-to-date data and realistic conditions. This includes assessing the harvested power in diverse scenarios: ground-to-ground links in both illuminated and dark lunar regions.

\end{itemize}

\label{section1-intro}

\section{System Model}

In our system model, we consider a ground-to-ground power transmission, with a range spanning from a few meters to tens of kilometers, as illustrated in Figure \ref{fig:scenario1}. The primary objective is to supply energy to rovers, vehicles, sensors, and habitats on the lunar surface. The system model assumes a clear line-of-sight between the power transmitter and receiver located near the lunar pole, with perfect beam alignment.

\subsection{Energy Harvesting and Beam Propagation Model}

\begin{figure}[h]
    \centering
    \includegraphics[width=0.5\textwidth]{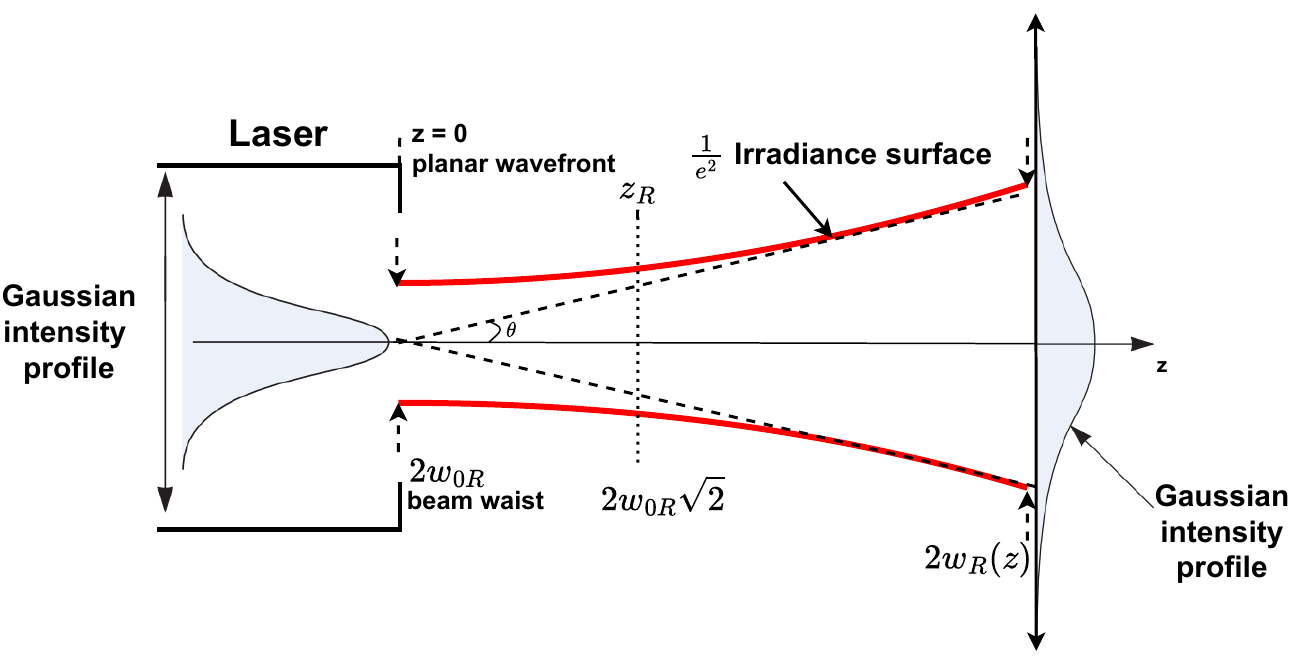}  
    \caption{Parameters of beam propagation, based on Gaussian theory.}
    \label{beam_prop}
\end{figure}

We aim to assess the amount of power that can be collected using realistic up-to-date data from the literature. The total harvested power, \( P_h \), is calculated by integrating the irradiance over the receiver area, taking into consideration the efficiency of the conversion processes from the laser power output to the electrical power received, and is expressed as

\begin{equation}
P_h = \eta_{\text{Rx}} \cdot \eta_{\text{dust}}(h,z) \cdot \int_A I(r, z) \, dA,
\end{equation}

\noindent where \( \eta_{\text{Rx}} \) denotes the efficiency of the receiver converting laser energy into electrical power, \( \eta_{\text{dust}} \) accounts for the attenuation of laser intensity due to scattering and absorption by lunar dust particles, $h$ represents the height of both the transmitter and receiver above the lunar surface and $z$ the transmission distance. The irradiance \( I(r, z) \) at a point on the receiver laser collector follows a two-dimensional Gaussian distribution, given by \cite{born_principles_1999}

\begin{equation}
I(r, z) = \frac{2P_t}{\pi w_R(z)^2} \cdot e^{-\frac{2r^2}{w_R(z)^2}},
\end{equation}

\noindent where \( P_t \) is the total transmitted power, \( r \) is the radial distance from the beam center, and \( w_R(z) \) is the beam radius at the propagation distance \( z \) from its waist (narrowest point of focus). This beam radius represents the distance at which the optical intensity drops to \( \frac{1}{e^2} \) (approximately 13.5\%) of the peak intensity at the plane \( z \) \cite{born_principles_1999}. These parameters describe the propagation of a Gaussian beam and are illustrated in Figure \ref{beam_prop}.\\

For a circular receiver laser collector with radius \( R \), an analytical expression for the harvested power can be derived as

\begin{equation}
P_h = \eta_{\text{cell}} \cdot \eta_{\text{dust}} \cdot P_t \cdot \left( 1 - e^{-\frac{2R^2}{w_R(z)^2}} \right).
\label{A}
\end{equation}

This equation relates the harvested power to the transmitted power \( P_t \), the receiver radius \( R \), and the beam radius \( w_R(z) \) at the distance \( z \) from the beam waist. \\

To quantitatively assess the harvested power and facilitate the Gaussian beam analysis, we consider the receiver laser collector radius matching the laser beam radius at the receiver location, such that \( R = w_R(z) \), meaning that the receiver is designed to capture approximately 87\% of the total power. This point is chosen because beyond it, the power gains become less and less significant compared to the required increase in receiver diameter. However, this implies that a substantial portion of the beam's power, around 13\%, spills over the edges of the receiver and is lost. Nevertheless, the reader should keep in mind that this is an arbitrary assumption to minimize size while capturing most of the power, and that the size can be chosen to meet for example other product specification points such as the full width at half maximum (FWHM), 95\%, $1/e^4$, and possibly others. In this specific case, equation (\ref{A}) becomes

\begin{equation}
    P_h = \eta_{\text{Rx}} \cdot \eta_{\text{dust}} \cdot P_t \cdot \left( 1 - e^{-2} \right).
    \label{B}
\end{equation}

The Rayleigh range, $z_R$, is the distance along the propagation axis of the Gaussian beam from its waist to the point where the cross-sectional area of the beam has doubled, corresponding to an increase in the beam's radius by a factor of $\sqrt{2}$, as defined in Figure \ref{beam_prop}. This distance represents the range over which the beam remains nearly collimated before diffraction causes significant divergence \cite{born_principles_1999}. Mathematically, it is defined as follows

\begin{equation}
z_R = \frac{\pi w_{0R}^2}{M^2 \lambda},
\label{Ray}
\end{equation}

\noindent where \(w_{0R}\) represents the beam waist of the real beam, \( M^2 \) represents the beam quality factor, and \( \lambda \) is the laser wavelength. The \(M^2\) parameter, mentioned in equation (\ref{Ray}), accounts for the deviation from the ideal Gaussian beam and is defined as follows \cite{siegman_defining_1993}

\begin{equation}
    M^2 = \frac{w_{0R} \theta_R}{w_0 \theta},
\end{equation}

\noindent with \(\theta_R\), representing the far-field divergence angle of the real beam. It is important to consider this factor in the modeling process, as for high-power or multi-mode laser operations, the \(M^2\) value can exceed 1. For example, currently available terrestrial high-power fiber lasers from IPG include single-mode lasers with power outputs up to 3 kW and beam quality factors of $M^2 < 1.1$ (YLR series). On the other hand, multi-mode lasers offer power outputs ranging from 1 kW to over 125 kW (YLS series), but with higher $M^2$ values, typically ranging from 6 to 50 \cite{IPG_high_power_fiber_lasers}.\\



 
The required receiver diameter, based on our \( \frac{1}{e^2} \) criterion defined earlier, is related to the transmitter diameter through Gaussian beam propagation. As stated above, our design choice for the receiver diameter is $d_{\text{Rx}} = 2R = 2w_R(z)$. Following the same idea, the transmitter diameter is $d_{\text{Tx}} = 2w_{0R}$. Therefore, based on the propagation equation \cite{born_principles_1999}, the relationship between both is expressed with the following equation 

\begin{equation}
    d_{\text{Rx}} = d_{\text{Tx}} \sqrt{1 + \left( \frac{z M^2 \lambda}{\pi \left( \frac{d_{\text{Tx}}}{2} \right)^2} \right)^2 }.
    \label{dx}
\end{equation} \\

This equation establishes the link between the transmitter and receiver diameters, based on our criterion of 87\% light collection, for a collimated laser beam at the transmitter output.\\

However, to explore the real theoretical limit of the system, a focusing system with a focal length $f$ is positioned at $z=0$, immediately after the laser output being perfectly collimated. This simple system can generalize any optical system with an emitted convergence (or divergence) represented by an effective focal length. Equation \ref{dx} becomes \cite{self1983}

\begin{equation}
d_{\text{Rx}} = 2 w_0' \sqrt{1 + \left(\frac{\lambda (z - z')}{\pi w_0'^2}\right)^2},
\label{focus}
\end{equation}

\noindent where \(w_0'\) represents the "image" waist (smallest spot size in front of the focusing system), and \(z'\) corresponds to the "image" waist position (\(z' > 0\)). These values can be calculated by applying the lens equation with Gaussian-optic correction \cite{self1983}, considering an "object" waist of size \(d_{\text{Tx}}/2\) located at \(z=0\). They are expressed as

\begin{equation}
z' = \frac{z_R^2 f}{f^2 + z_R^2},
\end{equation}

\noindent and

\begin{equation}
w_0' =  \frac{f d_{\text{Tx}}}{2\sqrt{f^2 + z_R^2}}.
\end{equation}

One should note that in the absence of a focusing system (\(f \to \infty\)), Equation \ref{focus}  reduces to Equation \ref{dx}. Under these conditions, the target transmission distance $z_{\text{target}}$ for a target $d_{Rx}$ taken at $z>z'$ is determined by the equation

\begin{equation}
\small
\begin{aligned}
    z_{\text{target}}(\alpha) = 
    \frac{\alpha}{\alpha^2 + 1} \cdot 
    \frac{\pi d_{\text{Tx}}}{4 \lambda M^2} 
    \Bigg[ 
        \sqrt{d_{\text{Rx}}^2 + 
        \left(d_{\text{Rx}}^2 - d_{\text{Tx}}^2\right) 
        \alpha^2} + d_{\text{Tx}} 
    \Bigg],
\end{aligned}
\label{eq11}
\end{equation}

\noindent where $\alpha = \frac{f}{z_R}$. By maximizing Equation \ref{eq11} through the selection of the focal length, the maximum possible transmission distance \(z_{\text{max}}\) as well as the optimal focal length \(f_{\text{opt}}\) can be determined, such that \(z_{\text{max}} = z_{\text{target}}(\alpha_{\text{opt}})\), where \(\alpha_{\text{opt}} = \frac{f_{\text{opt}}}{z_R}\). This maximum is calculated numerically for each combination of \(d_{\text{Tx}}\) and \(d_{\text{Rx}}\).

\subsection{Lunar Dust Attenuation Model}

\subsubsection{Radiative Transfer Theory}

The presence of lunar dust negatively affects laser transmission through scattering and absorption. The magnitude of this effect depends on factors such as the laser wavelength, the physical properties of the lunar dust, and its concentration. To quantify this attenuation, we aim to derive an expression for $\eta_{dust}$, using realistic, up-to-date data from the literature. The attenuation caused by the dust cloud can be modeled using the Radiative Transfer Equation (RTE). While solving the RTE is generally complex, our specific scenario allows for a set of assumptions that significantly simplify the problem.\\

The following hypotheses are adopted \cite{mishchenko_multiple_2006}:

\begin{enumerate}
    \item \textbf{Illumination by quasi-monochromatic light:}  
    Lasers are designed to emit light with an extremely narrow spectral bandwidth, effectively approximating quasi-monochromatic illumination.

    \item \textbf{Far-field particle interaction:}  
    Each particle in the scattering medium is assumed to be in the far-field zone relative to all other particles. This assumption is justified by the micrometer-scale size of the particles, the laser wavelength, and the relatively low particle density. 

    \item \textbf{Neglect of multiple scattering paths (Twersky approximation):}  
    The low particle density in the dust cloud ensures that the likelihood of a wave scattered by one particle returning to interact with the same particle is negligible. This simplifies calculations without significantly affecting accuracy, as the optical thickness is sufficiently low to validate this approximation.

    \item \textbf{Ergodicity hypothesis:}  
    The constant motion of particles in the dust cloud ensures that the time-averaged properties of the system are equivalent to an ensemble average over all possible particle positions and orientations. This assumption enables statistical simplifications in the modeling.

    \item \textbf{Statistical independence and uniform particle distribution along the optical path:}  
    Particles are considered statistically independent of one another. Furthermore, for the spatial scales relevant to the laser's interaction with the dust, the particle distribution is assumed to be uniform along the optical path. While the density varies with the subsolar angle, this assumption is valid for relatively short optical path distances.

    \item \textbf{Convex scattering medium:}  
    The scattering medium is assumed to be convex, which prevents re-entry of light once it leaves the medium. This assumption is consistent with the modeled scenario.
\end{enumerate}

Under these assumptions, the RTE simplifies to a form that links the intensity of the optical beam to the particle density and the extinction matrix \cite{mishchenko_multiple_2006}

\begin{equation}
\frac{d I_c (z)}{dz} = -n_0 K(\hat{z}) I_c(z)
\label{eq},
\end{equation}

\noindent where $I_c$ represents the coherent intensity vector, $z$ is the path length along the propagation direction $\hat{z}$, $n_0$ is the number density of particles in $particles/m^3$, and $K(\hat{z})$ is the extinction matrix.\\

For spherically symmetric, optically isotropic particles, we obtain

\begin{equation}
\eta_{\text{dust}} = \frac{I_c(z)}{I_c^{\text{inc}}} = e^{-\alpha_{\text{ext}} z},
\end{equation}

\noindent where $I_c^{\text{inc}}$ is the incident coherent intensity vector, $\alpha_{\text{ext}} = n_0 (C_{\text{ext}})$ is the extinction (attenuation) coefficient, and $\langle C_{\text{ext}} \rangle$ is the orientation-averaged extinction cross-section.

\subsubsection{T-matrix method}

\begin{figure}[h]
    \centering
    \includegraphics[width=0.6\textwidth]{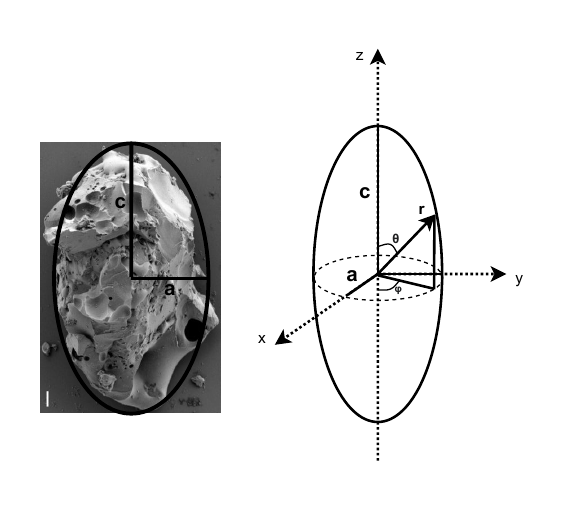}
    \caption{3D illustration of particle shape characterization by a prolate spheroid and its geometrical parameters along z, the axis of revolution. Picture credit : Chiaramonti Debay/NIST \cite{nist_mooning_2017}.}
    \label{Prolate spheroid picture}
\end{figure}

In order to determine $\langle C_{\text{ext}} \rangle$, a shape must be chosen to characterize a lunar dust particle. Yang Liu \textit{et al.} \cite{liu_characterization_2008} conducted a comprehensive study on the shape characteristics of lunar dust, focusing on parameters such as aspect ratio and complexity factor. Their analysis revealed that lunar dust particles exhibit irregular, angular shapes, distinguishing them from simpler geometric models such as spheres. They demonstrated that lunar dust can be well approximated by an elongated spheroid, also known as a prolate spheroid, as shown in Figure \ref{Prolate spheroid picture}. Using the least-squares method, they determined the average aspect ratio (defined as the ratio of the minor axis to the major axis) being around 0.7. While the prolate spheroid model may not capture every jagged edge or surface feature, it effectively represents the overall elongation, reflected in the aspect ratio, and provides a useful framework for studying the optical properties of lunar dust.\\

Specifically, the model uses the T-matrix method to simulate the scattering and absorption behavior of lunar dust particles \cite{mishchenko2002scattering}. Unlike Mie theory, which assumes spherical particles, the T-matrix method allows for a more accurate representation of their elongated shapes \cite{mishchenko2002scattering,quirantes_t-matrix_2005}. It is especially useful for determining the orientation-averaged extinction cross-section, which is needed to quantify the total amount of light blocked by a particle (both absorbed and scattered). More precisely, it enables efficient computation by computing the T-matrix for one orientation and applying rotational transformations to derive properties for other orientations.\\

A comprehensive explanation of the T-matrix method is provided in \cite{mishchenko2002scattering}, and the key concepts, notations, and specifics of the algorithm used are discussed in detail in \cite{somerville2016}. Rather than restating all this information, we will summarize the most relevant point. The orientation-averaged extinction cross-section $\langle C_{\text{ext}} \rangle$ is calculated using the T-matrix elements and can be computed as \cite{mishchenko2002scattering}

\begin{equation}
\langle C_{\text{ext}} \rangle = \frac{-2\pi}{k^2} \sum_{n=1}^{\infty} \sum_{m=0}^{n} (2 - \delta_{m,0}) \text{Re} \left( T_{nn|m}^{11} + T_{nn|m}^{22} \right),
\end{equation}

\noindent where $k= \frac{2\pi}{\lambda}$ is the wavenumber of the incident light in the surrounding medium, \(T_{nn|m}^{11}\) and \(T_{nn|m}^{22}\) are elements of the T-matrix that describe the scattering behavior of the particles for various \(n\) and \(m\).  The indices \(m\) and \(n\) correspond to the projected angular momentum along the z-axis and total angular momentum, respectively, with \( |m| \leq n \) and \(n = 1 \dots \infty\). The Kronecker delta \( \delta_{m0} \) ensures proper weighting for \(m = 0\) \cite{mishchenko2002scattering,somerville_new_2013}. We also have

\begin{equation}
\langle C_{\text{ext}} \rangle = \langle C_{\text{abs}} \rangle + \langle C_{\text{sca}} \rangle,
\end{equation}

\noindent where, $ \langle C_{\text{abs}} \rangle$ and $\langle C_{\text{sca}} \rangle$ designate the orientation-averaged absorption and scattering cross-section, respectively.\\

For further details, the detailed procedures of the T-matrix elements calculation, including the specific algorithms for matrix inversion and handling of numerical issues, the reader is referred to \cite{somerville2016}.

\subsubsection{Particle Size, Refractive Index, and Particle Density Distribution}

To account for the distribution of particle sizes, the rigorous approach would involve integrating the orientation-averaged extinction cross-section over all possible sizes, weighted by the normalized distribution of equivalent-sphere radii \( n(r) \). However, for electrostatically lofted particles, such a distribution remains unknown \cite{popel_lunar_2018}. Therefore, we define \( r_{\text{average}} \), referred to as "The average particle size", as the radius at which \( \langle C_{\text{ext}}(r) \rangle \) equals its mean value. Mathematically, this requires solving

\begin{equation}
\langle C_{\text{ext}}(r_{\text{average}}) \rangle = \langle C_{\text{ext}} \rangle_{\text{mean}},
\end{equation}

\noindent where the mean extinction cross-section \( \langle C_{\text{ext}} \rangle_{\text{mean}} \) over the given range \( r_{\text{min}} \) to \( r_{\text{max}} \) is

\begin{equation}
\langle C_{\text{ext}} \rangle_{\text{mean}} = \frac{1}{r_{\text{max}} - r_{\text{min}}} \int_{r_{\text{min}}}^{r_{\text{max}}} \langle C_{\text{ext}}(r) \rangle \, .
\end{equation}

Concerning the complex refractive index of lunar dust, it is expressed as

\begin{equation}
    \tilde{n} = \text{Re}(\tilde{n}) + i\text{Im}(\tilde{n}),
\end{equation}

\noindent where $\text{Re}(\tilde{n})$ describes the phase velocity of light as it propagates through the lunar dust medium. This real component primarily governs optical effects such as refraction, while the imaginary part quantifies the absorption of electromagnetic waves as they travel through the medium, indicating how much light is absorbed.\\

In the literature, the real part of the refractive index for lunar dust is typically reported to be between 1.58 and 1.78 \cite{refractiveindex1,refractiveindex2}. However, obtaining accurate values for the imaginary part is more difficult. For the purposes of this study, we approximate the refractive index by assuming that lunar dust particles are mostly composed of silicates. As a result, we base our refractive index values on those of "astronomical silicate" \cite{draine_optical_1984}. This approach simplifies the modeling process, given the variable composition of lunar dust and is a justified approximation as studies of lunar geology indicated that the regolith is rich in silicate minerals. Furthermore, the optical properties of astronomical silicates are well-documented, making them a suitable approximation for lunar dust in this context.\\

To determine the complex refractive index, we use the dielectric function, represented by the real part \( \epsilon_1 \) and the imaginary part \( \epsilon_2 \). Specifically, \( \epsilon_1 \), represents the material's ability to store electrical energy, while \( \epsilon_2 \) indicates the material's capacity to dissipate energy as heat, reflecting dielectric losses. Using these dielectric function values from the literature for a specified wavelength \cite{draine_optical_1984}, we calculate the real and imaginary parts of the complex refractive index with the following equations

\begin{equation}
\text{Re}(\tilde{n})(\lambda) = \sqrt{\frac{\sqrt{\epsilon_1^2 + \epsilon_2^2} + \epsilon_1}{2}},
\label{eq:A11}
\end{equation}
\begin{equation}
\text{Im}(\tilde{n})(\lambda) = \sqrt{\frac{\sqrt{\epsilon_1^2 + \epsilon_2^2} - \epsilon_1}{2}}.
\label{eq:A22}
\end{equation}

These expressions enable us to model the wavelength-dependent refractive index of lunar dust using the dielectric properties of astronomical silicate.\\

The dust grain density distribution above the lunar surface is determined based on the assumption that our scenario takes place in the polar regions under a high subsolar angle. For the ground-to-ground scenario, we focus on the near-surface dust density resulting from electrostatic levitation, neglecting the contribution from meteorite impacts that eject lunar dust into the exosphere, as the LADEE experiment has shown this effect to be negligible \cite{szalay_lunar_2016}.\\

The phenomenon of electrostatic dust lofting predominates in the illuminated regions of the Moon under solar radiation \cite{popel_lunar_2018}, \cite{golub_dusty_2012}. We do not analyze the effect in dark regions since \cite{naqbi_impact_2024} has shown this effect to be negligible. According to \cite{popel_lunar_2018}, the number density of charged dust grains in the near-surface layer of the Moon’s illuminated side is approximately \( 10^3 \, \text{cm}^{-3} \), though this value can vary depending on the subsolar angle. This high density is attributed to the abundance of photoelectrons, including those emitted directly from the surfaces of the dust grains themselves. The particles involved in this lofting process typically range in size from 0 to 0.2 micrometers.\\
 

The particle density distribution as a function of height above the lunar surface is given by the following equation, which is derived from \cite{popel_lunar_2018} using the least-squares fitting method

\begin{equation}
n(h) = -4.166 \times 10^8 \cdot \ln\left(\frac{h}{868}\right),
\end{equation}

\noindent valid for $0 \leq h \leq 868 \, \text{cm}$, where \( n(h) \) is in particles/m\(^3\) and \( h \) in centimeters. The parameters used in this distribution are sourced from \cite{popel_lunar_2018}, with a subsolar angle set at \( \theta = 87^\circ \), corresponding to the polar regions of the Moon. Additionally, the photoelectron density at the lunar surface is \( N_0 = 1.3 \times 10^2 \, \text{cm}^{-3} \), the thermal energy of the electrons is \( T_e = 1.3 \, \text{eV} \), and the quantum yield is based on data from the cited source. It should be noted that, although this distribution corresponds to a subsolar angle of 87 degrees, the density tends to vary only slightly with the subsolar angle \cite{popel_lunar_2018}.\\

\label{section2-Syst}

\section{Performance Evaluation}

The impact of lunar dust on the system’s performance is quantitatively assessed through simulation studies. Before calculating the harvested power, the extinction cross-sectional area of the lunar dust particles is determined. Next, the attenuation coefficient and the efficiency factor \( \eta_{\text{dust}} \) are evaluated in our ground-to-ground link on the lunar surface. The harvested power is then calculated under realistic conditions and assuming perfect tracking with no losses due to beam misalignment. Afterward, we address practical considerations and the validity of the chosen approach. Finally, we discuss design considerations for the transmitter and receiver apertures in order to maximize the transmission distance.

\subsection{Simulation Parameters}

The selection of the laser must optimize several criteria: high power, good beam quality, a wavelength that matches the photovoltaic cell, and high electrical-to-optical conversion efficiency. For these reasons, we choose a single-mode high-power fiber laser operating at a wavelength of 1064\,nm. Current high-power lasers are typically ytterbium-doped (Yb) fiber lasers with operating wavelength ranging from approximately 1007nm to 1070nm, depending on the design. Selecting 1064nm is acceptable and aligns with the peak efficiency of most PV receivers. Consequently, we set the wavelength at 1064\,nm for our simulation, meaning that the system must select a photovoltaic cell to match the laser wavelength, rather than the other way around. The transmitted power \(P_t\) is assumed to range between 1\,kW and 10\,kW. For this laser, we assume a reasonably low Beam Parameter Product of 2\,mm\(\cdot\)mrad, which corresponds to an \(M^2\) parameter of approximately 5.9, based on commercially available fiber laser performance.\\

Among available photovoltaic cells, indium-gallium-arsenide (InGaAs) are selected.
\(\text{In}_{0.24}\text{Ga}_{0.76}\text{As}\) cells are optimized for a bandgap around 1.08\,eV, which closely aligns with the laser wavelength of 1064\,nm. These cells have demonstrated a maximum conversion efficiency of 45.5\% under high-intensity illumination over small areas \cite{kalyuzhnyy_ingaas_2018}.\\

Based on the dielectric function data provided in \cite{draine_optical_1984} and the formulas (\ref{eq:A11}) and (\ref{eq:A22}), the complex refractive index for a wavelength of 1064 nm is determined to be \( n = 1.733 + i 0.05 \). The value of the real and imaginary part of the refractive index is within the range of values given in the literature for lunar dust, which solidifies the relevance of the chosen approach. Given space conditions, the refractive index of the medium surrounding the particle is set to 1.0. Concerning the aspect ratio, \( AR \), we define it as the ratio between the maximum and minimum distances from the origin, where \( c > a \) for elongated spheroids. Based on Yang Liu \textit{et al.} \cite{liu_characterization_2008}:

\begin{equation}
AR = \frac{r_{\text{max}}}{r_{\text{min}}} = \frac{c}{a}=1.428.
\end{equation}
\noindent Table \ref{tab:parameters1} summarizes the parameters for our considered scenario.

\begin{table}[htb]
    \caption{Summary of Parameters}
    \centering
    \begin{tabular}{|l|c|}
    \hline\rowcolor[gray]{0.8}\color{black}
    \bf{Parameter} & \multicolumn{1}{c|}{\bf{Value}} \\\hline
        Transmission Range & up to 20 km \\\hline
        Transmitted Power & 1 to 10 kW \\\hline
        Particle Size Range & up to 200 nm \\\hline
        Particle Complex Refractive Index & \(1.733 + i\,0.05\) \\\hline
        Particle Average Aspect Ratio & 1.4286 \\\hline
        Medium Refractive Index & 1.0 \\\hline
        Laser Wavelength \((\lambda)\) & 1064 nm \\\hline
        $M^2$ & 5.9 \\\hline
        EHCE (\%) & 45.5\% \cite{kalyuzhnyy_ingaas_2018} \\\hline
    \end{tabular}
    \label{tab:parameters1}
\end{table}

\subsection{Results and discussions}

\subsubsection{Dust-related Attenuation}

First of all, simulations were carried out to assess the influence of lunar dust on laser power transmission for our ground-to-ground link scenario. Then, the results of these simulations were used to determine the power harvested by a receiver, considering varying transmit powers and distances \( z \) within our proposed system model.\\

\begin{figure}[htb]
    \centering
    \includegraphics[width=0.5\textwidth]{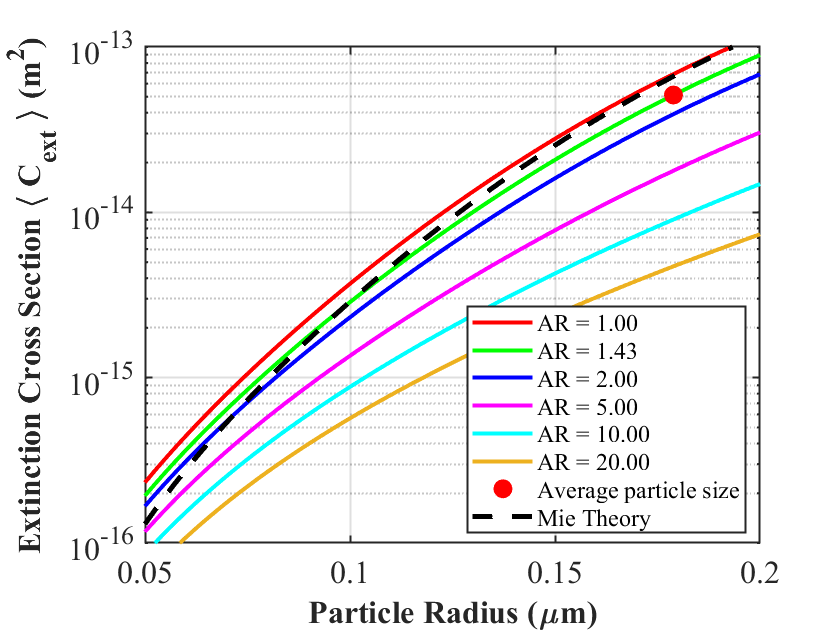}
    \caption{Averaged extinction cross-section as a function of particle radius for various particle aspect ratios, with the average particle size marker used.}
    \label{fig:C_ext_vs_R}
\end{figure}

\begin{figure}[!htb]
    \centering
    \includegraphics[width=0.5\textwidth]{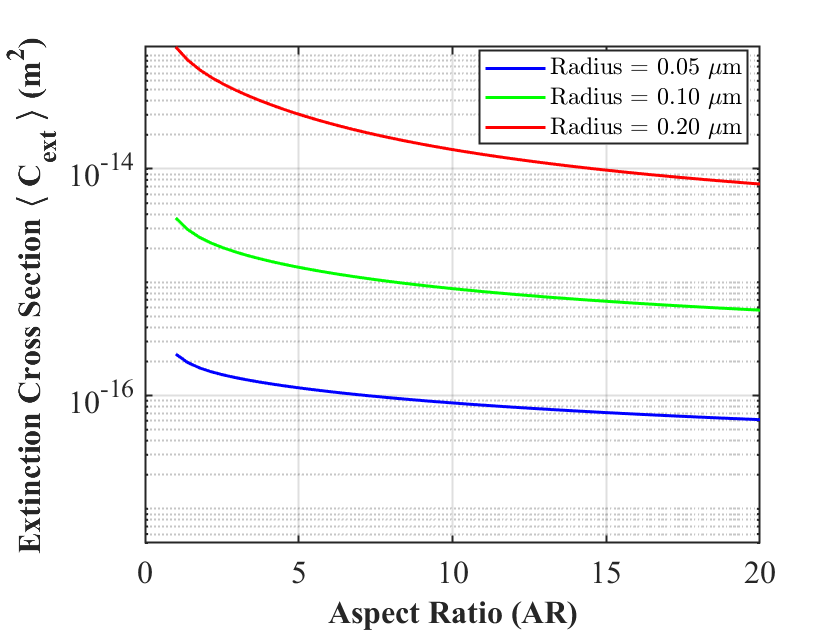}
    \caption{Averaged extinction cross-section as a function of particle aspect ratio for different particle radii.}
    \label{fig:C_ext_vs_AR}
\end{figure}

The average extinction cross-section was determined using the methodology outlined in Section 2, in conjunction with the data specified in Table \ref{tab:parameters1}. The results are presented as a function of particle radius and aspect ratio, and are illustrated in Figures \ref{fig:C_ext_vs_R} and \ref{fig:C_ext_vs_AR}.\\

One of the key observations is the significant influence of particle size on the extinction cross-section. Although it might be expected that larger particles show an increased ability to absorb or scatter laser light, this increase appears to follow an exponential trend, clearly demonstrating that particle size plays a crucial role. Importantly, the average particle size defined in the system model section is identified in Figure \ref{fig:C_ext_vs_R} for the considered average aspect ratio, dividing the area equally and providing a representative marker for the overall behavior of the extinction cross-section over the particle range. It is also worth noting that the red curve corresponds to the standard Mie theory \cite{craig_f_bohren_donald_r_huffman_absorption_1998}, to which our T-matrix calculation closely aligns within the region of interest.\\


Furthermore, the curves corresponding to different aspect ratios reveal that as the particle aspect ratio increases, the extinction cross-section decreases. This suggests that more elongated particles tend to scatter or absorb less laser light on average. Figure \ref{fig:C_ext_vs_AR} further highlights this influence.\\

\begin{figure}[h]
    \centering
    \includegraphics[width=0.5\textwidth]{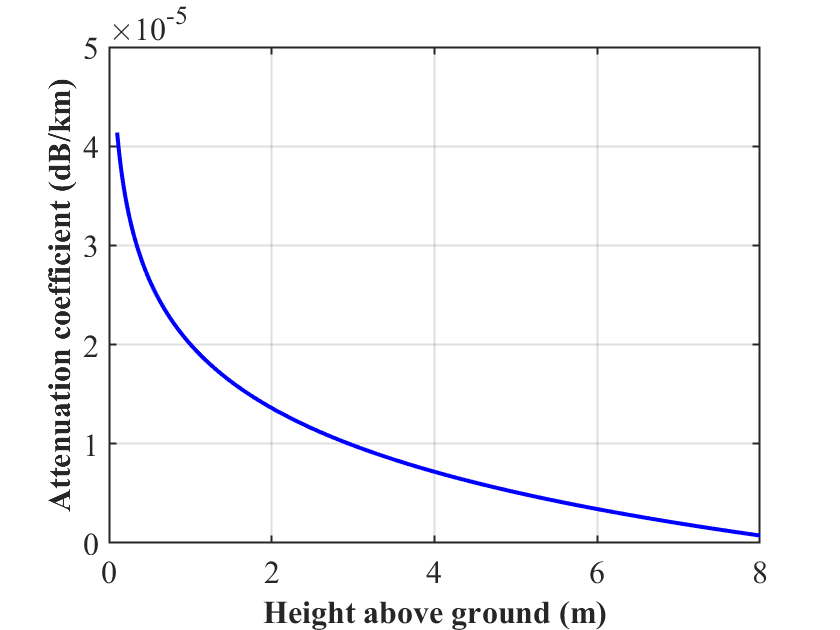}
    \caption{Attenuation coefficient as a function of height above the lunar surface.}
    \label{fig:att_coef}
\end{figure}

An important feature to present is the attenuation coefficient as a function of height, as illustrated in Figure \ref{fig:att_coef}. It is worth noting that, in our case, the attenuation coefficient mirrors the density, given that our $\langle C_{\text{ext}} \rangle$ has been defined based on a mean particle size and aspect ratio.\\

\begin{figure}[h]
    \centering
    \includegraphics[width=0.5\textwidth]{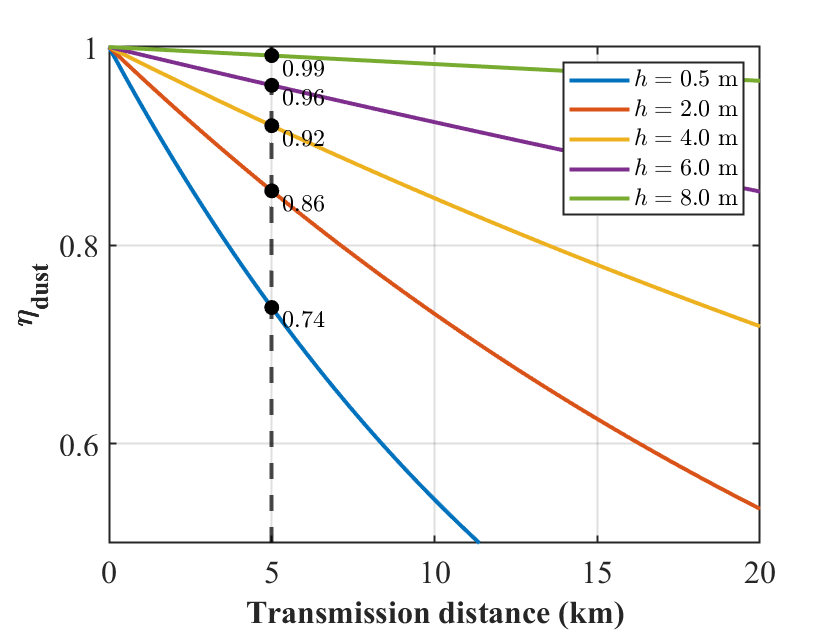}
    \caption{Dust transmission efficiency ($\eta_{\text{dust}}$) as a function of transmission distance for different transmission heights.}
    \label{fig:eta_dust_ground_link}
\end{figure}

We then analyze the dust-related efficiency as a function of transmission distance for different heights. Figure \ref{fig:eta_dust_ground_link} clearly demonstrates that lunar dust can have a significant impact on transmission efficiency near the surface. Naturally, the greater the transmission distance, the more lunar dust interferes with the laser beam, through absorption and scattering mechanisms. For a realistic transmission distance of 5 km, the losses are estimated.\\

\begin{figure}[h]
    \centering
    \includegraphics[width=0.5\textwidth]{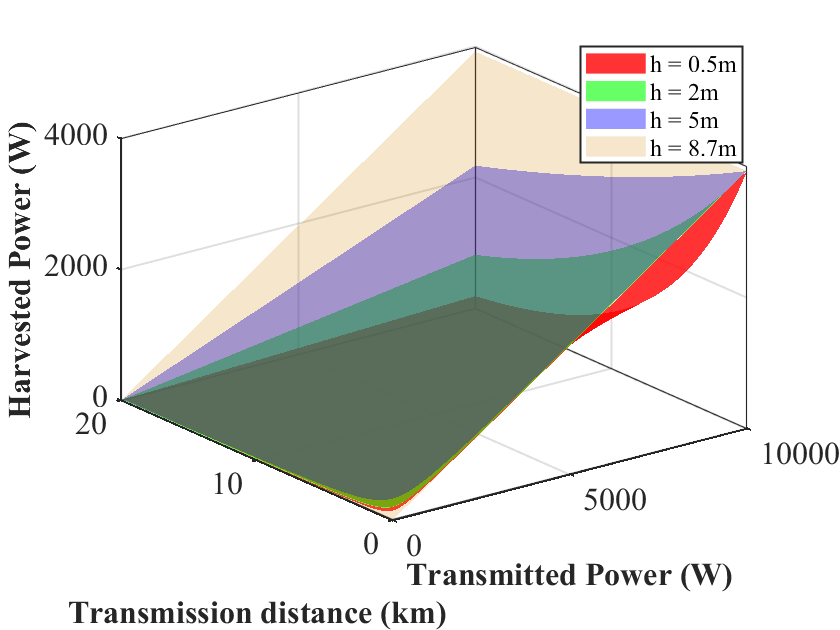}
    \caption{Harvested power as a function of transmitted power and transmission distance for different transmission heights in a ground-to-ground link, based on equation (\ref{B}).}
    \label{fig:Ph_3D_plot}
\end{figure}

Figure \ref{fig:Ph_3D_plot} illustrates the harvested power as a function of transmission distance and transmitted power at various heights, based on equation \ref{B} which is independent of the transmitter and receiver size. The graph shows that close to the lunar surface, due to the influence of lunar dust, the amount of harvested power can be significantly reduced. To minimize these losses, it becomes advantageous to position transmitters and receivers at a certain height, where the dust's impact becomes negligible. In this case, the primary source of energy loss is due to the conversion efficiency of the collector panel, which converts laser energy into electrical energy, as well as the attenuation from the \(\frac{1}{e^2}\) factor of the Gaussian beam, in our case. For instance, if we aim to transmit 5 kW over a distance of 15 km, the harvested power would be 774 W at 0.5 meters, 1211 W at 2 meters, 1629 W at 5 meters, and 1947 W at 9 meters above the surface, where the influence of lunar dust becomes negligible.\\

\subsubsection{Limitations}

In reality, the transmission distances which extend up to 20 km, shown in Figure \ref{fig:Ph_3D_plot}, may be challenging to achieve near the surface due to the curvature of the Moon and its smaller size compared to Earth. Such line-of-sight distances are, therefore, only realistic at higher altitudes to overcome the horizon, as demonstrated in Figure \ref{fig:Tx_Rx_range}. The equation that describes this (see the derivation in the appendix) is the following

\begin{equation}
    LoS_{range}(h_t, h_r) = \sqrt{h_t^2 + 2h_tR} + \sqrt{h_r^2 + 2h_rR}
\end{equation}

\noindent where \( h_t \) represents the height of the transmitter, \( h_r \) represents the height of the receiver, and \( R \) is the radius of the Moon (considered perfectly spherical). In our system model shown in Figure \ref{fig:scenario1}, we assume $h_t=h_r$ for the simplicity of the analysis, and the consistency of our model. Indeed, in our lunar dust attenuation model, we assume a flat lunar surface between the transmitter and receiver, and based on Figure \ref{fig:Tx_Rx_range}, this assumption holds true for sufficiently small distances, typically below 5 km. Beyond this, the accuracy of our lunar dust attenuation model tends to decrease, as the dust density tends to vary along the LoS. It should also be noted that near-surface beam propagation at lower altitudes can be affected by the lunar surface topology, including crater walls, hills, and ridges, which can affect the optical propagation along the lunar surface depending on the specific location where OPB is taking place. However, our simplified scenario does not take these topographical variations into account, as it assumes a flat lunar surface. Incorporating topographical details would require a more complex model, introducing the need to know the specific topology of the analyzed location. \\

\begin{figure}[!htb]
    \centering
    \includegraphics[width=0.5\textwidth]{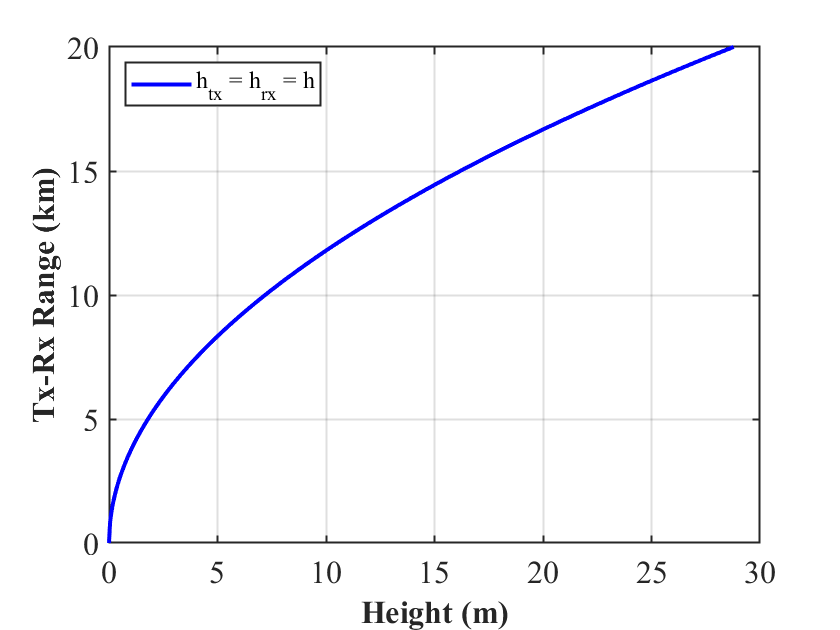}
    \caption{Possible transmitter-receiver range on the Moon as a function of transmission height.}
    \label{fig:Tx_Rx_range}
\end{figure}

In general, the applicability and accuracy of our ground-to-ground model are limited to short transmission distances, typically a few kilometers at most. Several factors contribute to this limitation. Firstly, as the transmission distance increases, the curvature of the Moon becomes more significant, making the optical path more complex. Secondly, multiple scattering effects tend to become more pronounced over longer distances, further affecting the attenuation of the transmitted laser light. Lastly, while the variation in particle density due to changes in the subsolar angle is expected to have a minor effect, it may still contribute to slight variations in the model’s predictions. These factors collectively restrict the validity of the model to shorter transmission ranges.\\

\subsubsection{Maximizing Transmission Distance}

\begin{figure}[!htb]
    \centering
    \includegraphics[width=0.5\textwidth]{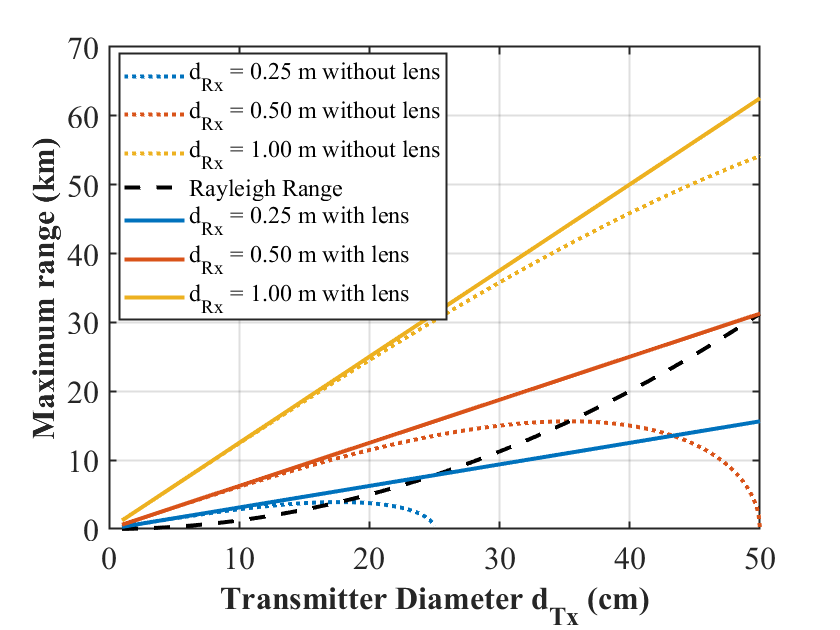}
    \caption{Maximum transmission distance as a function of transmitter diameter for different receiver diameter, to maintain 87\% of the collected light, without lens (collimated) and with optimal lens.}
    \label{compact}
\end{figure}

When designing such a system for space applications, it is essential to ensure that the aperture sizes are compact while enabling long transmission distances to effectively reach the target. By focusing solely on the transmitter and receiver sizes and optimizing for maximum transmission distance, the optimal design will depend on the relative size of the receiver compared to the transmitter, as illustrated in Figure \ref{compact}. Without a lens (focusing system), the laser is launched in a perfectly collimated manner, and as long as the target diameter is larger than that of the transmitter, a solution can be achieved, as shown by the dotted lines in Figure \ref{compact}. In this case, the maximum range is the Rayleigh range determined by the transmitter aperture diameter. Introducing an optimal focusing lens further enhances the transmission distance and provides a solution even when the transmitter diameter exceeds the target diameter, as the lens allows the beam to be focused. For distances far beyond the Rayleigh range, however, the optimal configuration with a lens converges with that without a lens, as diffraction ultimately limits how tightly a beam can be focused.
\\

\begin{figure}[!htb]
    \centering
    \includegraphics[width=0.5\textwidth]{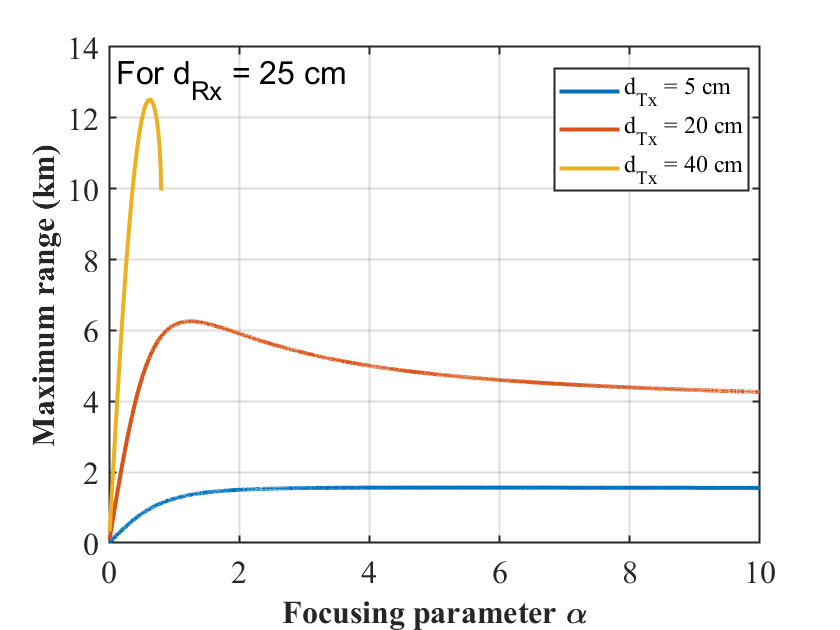}
    \caption{Maximum transmission distance as a function of the focusing parameter for three different transmitter diameters, while fixing the receiver diameter to 25 cm.}
    \label{curves_graph}
\end{figure}

Figure \ref{curves_graph} gives us more insight about the behavior of the system. The three curves correspond to the three possible scenarios. When we are far from the Rayleigh range (blue curve), diffraction dominates, and adding a focusing system with extremely long focal lengths does not improve performance, as the beam's natural divergence cannot be overcome. When we are near the Rayleigh range (red curve), adding an optimized focusing system slightly extends the maximum range and shifts the optimal point, improving system performance. When we are within the Rayleigh range (yellow curve), a focusing system is essential to efficiently reach the target, but not all focal lengths provide a solution due to limits in beam concentration and spot size.\\

\label{section3-Eval}

\section{Conclusions}

In this study, we analyzed the performance of a ground-to-ground optical power transmission system on the lunar surface. By employing radiative transfer theory and the T-matrix method, we modeled the scattering and absorption effects of lunar dust particles, considering their elongated shapes, sizes, and refractive indices, to assess the resulting optical power attenuation. Our simulations demonstrated that electrostatically lofted lunar dust (LLD) can significantly attenuate optical power transmission near the lunar surface in illuminated regions. Additionally, we observed that multiple kilowatts of energy harvesting are achievable with current technologies and concluded that in these regions, the devices had to be placed at a certain height to avoid dust-related losses and maximize energy harvesting performance.\\

Furthermore, we investigated designs that maximize transmission distance and concluded that incorporating a focusing system with the appropriate focal length optimizes transmission distance when the receiver is near or within the Rayleigh range. However, when the distance is far beyond the Rayleigh range, focusing becomes unnecessary, and launching the beam in a perfectly collimated manner is the optimal solution. In summary, this work provides valuable insights into optimizing OPB systems in dusty lunar environments, contributing to reliable energy solutions for future lunar exploration missions.

\label{section4-Conclusion}

\label{section5-Appendix}

\section*{ACKNOWLEDGMENT}

This work was supported  by the Tier 1 Canada Research Chair program.

\section*{REFERENCES AND FOOTNOTES}

\printbibliography

@article{lin_return_2024,
  title = {Return to the {Moon}: {New} perspectives on lunar exploration},
  author = {Lin, Yangting and Yang, Wei and Zhang, Hui and Hui, Hejiu and Hu, Sen and Xiao, Long and Liu, Jianzhong and Xiao, Zhiyong and Yue, Zongyu and Zhang, Jinhai and Liu, Yang and Yang, Jing and Lin, Honglei and Zhang, Aicheng and Guo, Dijun and Gou, Sheng and Xu, Lin and He, Yuyang and Zhang, Xianguo and Qin, Liping and Ling, Zongcheng and Li, Xiongyao and Du, Aimin and He, Huaiyu and Zhang, Peng and Cao, Jinbin and Li, Xianhua},
  journal = {Science Bulletin},
  year = {2024},
  volume = {69},
  number = {13},
  pages = {2136--2148},
  month = jul,
  doi = {10.1016/j.scib.2024.04.051},
  shorttitle = {Return to the Moon},
  urldate = {2024-09-24},
  langid = {english}
}

@article{popel_lunar_2018,
  author = {Popel, S.I. and Zelenyi, L.M. and Golub', A.P. and Dubinskii, A.Yu.},
  title = {Lunar dust and dusty plasmas: Recent developments, advances, and unsolved problems},
  journal = {Planetary and Space Science},
  year = {2018},
  volume = {156},
  pages = {71--84},
  month = jul,
  doi = {10.1016/j.pss.2018.02.010},
  shorttitle = {Lunar dust and dusty plasmas},
  urldate = {2024-09-24},
  langid = {english}
}

@inproceedings{gage_laser_2011,
  author = {Nugent, Jr., Thomas J. and Kare, Jordin T.},
  title = {Laser power beaming for defense and security applications},
  booktitle = {{SPIE} Defense, Security, and Sensing},
  editor = {Gage, Douglas W. and Shoemaker, Charles M. and Karlsen, Robert E. and Gerhart, Grant R.},
  year = {2011},
  pages = {804514},
  month = may,
  doi = {10.1117/12.886169},
  location = {Orlando, Florida, United States},
  eventtitle = {{SPIE} Defense, Security, and Sensing},
  langid = {english}
}

@inproceedings{olsen1991,
  author    = {Olsen, L. C. and Dunham, G. and Huber, D. A. and Addis, F. W. and Anheier, N. and Coomes, E. P.},
  title     = {GaAs solar cells for laser power beaming},
  booktitle = {NASA. Lewis Research Center, Space Photovoltaic Research and Technology Conference},
  year      = {1991},
  month     = {August}
}

@inproceedings{vanriesen2002,
  author    = {Van Riesen, S. and Schubert, U. and Bett, A. W.},
  title     = {GaAs photovoltaic cells for laser power beaming at high power densities},
  booktitle = {Proc. 17th Eur. PV Solar Energy Conf},
  pages     = {18--21},
  year      = {2001},
  month     = {October}
}

@inproceedings{raible2008,
author = {Raible, D. E.},
title = {High-intensity laser power beaming for long-range wireless power transmission},
booktitle = {Proceedings of SPIE},
volume = {7005},
pages = {70051F},
year = {2008}
}

@online{NASA_LaserMotive_2009,
  author       = {{NASA Administrator}},
  title        = {LaserMotive Wins \$900,000 from NASA in Space Elevator Games},
  organization = {NASA},
  year         = {2009},
  url          = {http://www.nasa.gov/centers/dryden/status_reports/power_beam.html}
}

@misc{NRL2019,
  author       = {{U.S. Naval Research Laboratory}},
  title        = {Researchers Transmit Energy with Laser in Historic Power Beaming Demonstration},
  howpublished = {\url{https://www.nrl.navy.mil/Media/News/Article/2504007/researchers-transmit-energy-with-laser-in-historic-power-beaming-demonstration}},
  year         = {2019},
}

@article{rennilson1974surveyor,
  author    = {Rennilson, J. J. and Criswell, D. R.},
  title     = {Surveyor observations of lunar horizon-glow},
  journal   = {The Moon},
  volume    = {10},
  number    = {2},
  pages     = {121--142},
  year      = {1974}
}

@book{mccoy1974lunar,
  author    = {McCoy, J. E. and Criswell, D. R.},
  title     = {Evidence for a high altitude distribution of lunar dust},
  year      = {1974},
  publisher = {Pergamon Press}
}

@article{horanyi2015dust,
  author    = {Hor{\'a}nyi, M. and Szalay, J. R. and Kempf, S. and Schmidt, J. and Gr{\"u}n, E. and Srama, R. and Sternovsky, Z.},
  title     = {A permanent, asymmetric dust cloud around the Moon},
  journal   = {Nature},
  volume    = {522},
  number    = {7556},
  pages     = {324--326},
  year      = {2015}
}

@inproceedings{stavnes_analysis_1992,
  author = {Stavnes, Mark W.},
  title = {An Analysis of Power Beaming for the Moon and Mars},
  booktitle = {27th Intersociety Energy Conversion Engineering Conference},
  year = {1992},
  month = aug,
  doi = {10.4271/929436},
  urldate = {2024-09-24},
  langid = {english}
}

@inproceedings{marcinkowski_lunar_2023,
  author = {Marcinkowski, Adam and Carrio, Luis and Hilliard, Sommer and Edwards, Christine and Elhawary, Alya and Clem, Dylan and Blood, Mikaela and May, Lisa and Cichan, Timothy},
  title = {Lunar Surface Power Architecture Concepts},
  booktitle = {2023 {IEEE} Aerospace Conference},
  year = {2023},
  pages = {1--19},
  month = mar,
  doi = {10.1109/AERO55745.2023.10115621},
  urldate = {2024-09-24},
  eventtitle = {2023 {IEEE} Aerospace Conference}
}

@incollection{craig_f_bohren_donald_r_huffman_absorption_1998,
  author = {Bohren, Craig F. and Huffman, Donald R.},
  title = {Absorption and Scattering by a Sphere},
  booktitle = {Absorption and Scattering of Light by Small Particles},
  pages = {82--129},
  publisher = {John Wiley \& Sons, Ltd},
  year = {1998},
  doi = {10.1002/9783527618156.ch4},
  isbn = {978-3-527-61815-6},
  urldate = {2023-10-05},
  langid = {english}
}

@article{liu_characterization_2008,
  author = {Liu, Yang and Park, Jaesung and Schnare, Darren and Hill, Eddy and Taylor, Lawrence A.},
  title = {Characterization of Lunar Dust for Toxicological Studies. {II}: Texture and Shape Characteristics},
  journal = {Journal of Aerospace Engineering},
  shortjournal = {J. Aerosp. Eng.},
  volume = {21},
  number = {4},
  pages = {272--279},
  year = {2008},
  month = {October},
  doi = {10.1061/(ASCE)0893-1321(2008)21:4(272)},
  issn = {0893-1321, 1943-5525},
  urldate = {2023-10-05},
  langid = {english}
}

@article{refractiveindex1,
  author = {Fearnside, Andrew and Masding, Philip and Hooker, Chris},
  title = {Polarimetry of moonlight: A new method for determining the refractive index of the lunar regolith},
  journal = {Icarus},
  volume = {268},
  pages = {156--171},
  year = {2016},
  doi = {10.1016/j.icarus.2015.11.038},
  urldate = {2023-10-05},
  month = apr,
  shortjournal = {Icarus}
}

@article{refractiveindex2,
  author = {Goguen, Jay D. and Stone, Thomas C. and Kieffer, Hugh H. and Buratti, Bonnie J.},
  title = {A new look at photometry of the Moon},
  journal = {Icarus},
  volume = {208},
  number = {2},
  pages = {548--557},
  year = {2010},
  doi = {10.1016/j.icarus.2010.03.025},
  urldate = {2023-10-05},
  month = aug,
  shortjournal = {Icarus}
}

@article{draine_optical_1984,
  author = {Draine, B. T. and Lee, H. M.},
  title = {Optical properties of interstellar graphite and silicate grains},
  journal = {The Astrophysical Journal},
  volume = {285},
  pages = {89},
  year = {1984},
  doi = {10.1086/162480},
  urldate = {2023-10-05},
  month = oct,
  shortjournal = {{ApJ}}
}

@inproceedings{tisdale_design_nodate,
  author    = {Tisdale, M. and Dulá, I. and Pabon Madrid, L. and Verkhovodova, P. and Pénot, J. and Coimbra, K. and Chung, S. J.},
  title     = {Design of a Modular and Orientable Electrodynamic Shield for Lunar Dust Mitigation},
  booktitle = {AIAA SCITECH Forum},
  year      = {2022},
  pages     = {2623}
}

@online{nist_mooning_2017,
  author = {Ann Chiaramonti Debay},
  title = {Mooning Over Measurement},
  year = {2017},
  url = {https://www.nist.gov/blogs/taking-measure/mooning-over-measurement},
  urldate = {2024-10-03},
  organization = {NIST - National Institute of Standards and Technology}
}

@article{zelenyi_dusty_2020,
  author    = {Zelenyi, L. M. and Popel, S. I. and Zakharov, A. V.},
  title     = {Dusty plasma at the Moon. Challenges of modeling and measurements},
  journal   = {Plasma Physics Reports},
  year      = {2020},
  volume    = {46},
  pages     = {527--540}
}

@article{horanyi_lunar_2024,
  author    = {Horányi, M. and Szalay, J. R. and Wang, X.},
  title     = {The lunar dust environment: concerns for Moon-based astronomy},
  journal   = {Philosophical Transactions of the Royal Society A},
  year      = {2024},
  volume    = {382},
  number    = {2271},
  pages     = {20230075}
}

@inproceedings{naqbi_impact_2024,
  author    = {Naqbi, M. and Loranger, S. and Karabulut Kurt, G.},
  title     = {Impact of Lunar Dust on Free Space Optical (FSO) Energy Harvesting},
  booktitle = {2024 IEEE Aerospace Conference},
  year      = {2024},
  month     = {March},
  pages     = {1--9},
  publisher = {IEEE}
}

@article{szalay_impact_2019,
  author    = {Szalay, J. R. and Pokorný, P. and Sternovsky, Z. and Kupiha, Z. and Poppe, A. R. and Horanyi, M.},
  title     = {Impact ejecta and gardening in the lunar polar regions},
  journal   = {Journal of Geophysical Research: Planets},
  year      = {2019},
  volume    = {124},
  number    = {1},
  pages     = {143--154}
}

@article{somerville2016,
  title = {SMARTIES: User-friendly codes for fast and accurate calculations of light scattering by spheroids},
  author = {Somerville, W. R. C. and Auguié, B. and Le Ru, E. C.},
  journal = {Journal of Quantitative Spectroscopy and Radiative Transfer},
  year = {2016},
  volume = {174},
  pages = {39--55},
  month = {}
}

@incollection{siegman_defining_1993,
  author = {Siegman, A. E.},
  title = {Defining and {Measuring} {Laser} {Beam} {Quality}},
  booktitle = {Solid {State} {Lasers}: {New} {Developments} and {Applications}},
  editor = {Inguscio, Massimo and Wallenstein, Richard},
  pages = {13--28},
  year = {1993},
  address = {Boston, MA},
  publisher = {Springer US},
  doi = {10.1007/978-1-4615-2998-9_2},
  language = {en}
}

@book{born_principles_1999,
  author = {Born, Max and Wolf, Emil},
  title = {Principles of {Optics}: {Electromagnetic} {Theory} of {Propagation}, {Interference} and {Diffraction} of {Light}},
  edition = {7},
  year = {1999},
  publisher = {Cambridge University Press},
  address = {Cambridge},
  doi = {10.1017/CBO9781139644181}
}

@misc{IPG_high_power_fiber_lasers,
  title = {High Power Fiber Lasers},
  author = {IPG Photonics},
  year = {2024},
  howpublished = {\url{https://www.ipgphotonics.com/products/lasers/industrial-cw-fiber-lasers/high-power-fiber-lasers}},
}

@book{mishchenko2002scattering,
  title = {Scattering, Absorption, and Emission of Light by Small Particles},
  author = {Michael I. Mishchenko and Larry D. Travis and Andrew A. Lacis},
  year = {2002},
  publisher = {Cambridge University Press},
  address = {Cambridge, UK},
  isbn = {0 521 78252 X},
  url = {http://www.giss.nasa.gov/~crmim/books.html}
}

@article{somerville_new_2013,
  author = {Somerville, W.R.C. and Auguié, B. and Le Ru, E.C.},
  title = {A new numerically stable implementation of the {T}-matrix method for electromagnetic scattering by spheroidal particles},
  journal = {Journal of Quantitative Spectroscopy and Radiative Transfer},
  volume = {123},
  pages = {153--168},
  year = {2013},
  doi = {10.1016/j.jqsrt.2013.01.023},
  month = {jul},
}

@article{quirantes_t-matrix_2005,
  author = {Quirantes, Arturo},
  title = {A {T}-matrix method and computer code for randomly oriented, axially symmetric coated scatterers},
  journal = {Journal of Quantitative Spectroscopy and Radiative Transfer},
  volume = {92},
  pages = {373--381},
  year = {2005},
  doi = {10.1016/j.jqsrt.2004.08.004},
  month = {may},
}

@book{mishchenko_multiple_2006,
  author = {Mishchenko, Michael I. and Travis, Larry D. and Lacis, Andrew and Lacis, Andrew A.},
  title = {Multiple scattering of light by particles: radiative transfer and coherent backscattering},
  publisher = {Cambridge Univ. Press},
  address = {Cambridge},
  year = {2006},
  edition = {1. publ},
  isbn = {978-0-521-83490-2},
}

@article{golub_dusty_2012,
  title = {Dusty plasma system in the surface layer of the illuminated part of the moon},
  volume = {95},
  issn = {1090-6487},
  doi = {10.1134/S0021364012040054},
  language = {en},
  number = {4},
  urldate = {2024-09-26},
  journal = {JETP Letters},
  author = {Golub’, A. P. and Dol’nikov, G. G. and Zakharov, A. V. and Zelenyi, L. M. and Izvekova, Yu. N. and Kopnin, S. I. and Popel, S. I.},
  month = {apr},
  year = {2012},
  pages = {182--187},
}

@article{szalay_lunar_2016,
  title = {Lunar meteoritic gardening rate derived from in situ {LADEE}/{LDEX} measurements},
  volume = {43},
  issn = {1944-8007},
  doi = {10.1002/2016GL069148},
  language = {en},
  number = {10},
  urldate = {2024-08-17},
  journal = {Geophysical Research Letters},
  author = {Szalay, Jamey R. and Horányi, Mihály},
  year = {2016},
  pages = {4893--4898},
}

@inproceedings{kalyuzhnyy_ingaas_2018,
  author    = {Kalyuzhnyy, N. A. and Emelyanov, V. M. and Mintairov, S. A. and Shvarts, M. Z.},
  title     = {InGaAs metamorphic laser ($\lambda$ = 1064 nm) power converters with over 44\% efficiency},
  booktitle = {AIP Conference Proceedings},
  year      = {2018},
  month     = {September},
  volume    = {2012},
  number    = {1},
  publisher = {AIP Publishing}
}

@article{colwell_lunar_2007,
  title = {Lunar surface: Dust dynamics and regolith mechanics},
  author = {Colwell, J. E. and Batiste, S. and Horányi, M. and Robertson, S. and Sture, S.},
  journal = {Reviews of Geophysics},
  volume = {45},
  number = {2},
  year = {2007}
}

@article{self1983,
  author    = {Self, S. A.},
  title     = {Focusing of spherical Gaussian beams},
  journal   = {Applied Optics},
  volume    = {22},
  number    = {5},
  pages     = {658--661},
  year      = {1983}
}






\end{document}